Coreless Terrestrial Exoplanets

Linda T. Elkins-Tanton and Sara Seager
Dept. Earth, Atmospheric, and Planetary Sciences
MIT
77 Massachusetts Ave
Cambridge MA 02139

Linda T. Elkins-Tanton: Building 54-824, ltelkins@mit.edu

Sara Seager (co-appointed to Dept. of Physics): Building 54-1626, seager@mit.edu

**Abstract**
Differentiation in terrestrial planets is expected to include the formation of a metallic iron core. We predict the existence of terrestrial planets that have differentiated but have no metallic core—planets that are effectively a giant silicate mantle. We discuss two paths to forming a coreless terrestrial planet, whereby the oxidation state during planetary accretion and solidification will determine the size or existence of any metallic core. Under this hypothesis, any metallic iron in the bulk accreting material is oxidized by water, binding the iron in the form of iron oxide into the silicate minerals of the planetary mantle. The existence of such silicate planets has consequences for interpreting the compositions and interior density structures of exoplanets based on their mass and radius measurements.

Subject headings;
accretion
planets and satellites: formation
solar system: formation

# 1. INTRODUCTION

Understanding the interior composition of exoplanets based on their mass and radius measurements is a growing field of study. While over 30 transiting exoplanets have measured masses and radii, all are between Jupiter (11.2 $R_\oplus$) and Neptune ($R_\oplus$) size. The anticipation of a large number of lower-mass, solid exoplanets to be discovered in the next few years is increasing, with radial velocity observations pushing to lower masses, and ground- and space-based surveys in operation or being readied for launch. These include the space-based transit surveys COROT (launched December 2007; Baglin, 2003) and Kepler (launch date 2009; Borucki et al. 2003), a ground-based transit search around low-mass stars MEarth (Nutzman & Charbonneau 2007); and ground-based radial velocity searches of low-mass stars (e.g., Bonfils et al. 2007, Endl et al. 2007, Butler et al. 2004) for which followup photometry may reveal transits.

For a solid planet with a given mass and radius, planet interior models aim to provide the mass fraction of planet building materials, typically assumed to be iron, silicate, and water. Even with the additional assumption of planetary differentiation, the actual composition of the interior will remain ambiguous (Valencia et al. 2007, Li and Seager 2008). This is due to the very different densities of iron, silicate perovskite, and water ice VII (zero pressure densities of 8.1 g cm$^{-3}$, ~4.3 g cm$^{-3}$, and 1.46 g cm$^{-3}$ respectively). For example, a planet with a relatively large fraction of silicate could have the same mass and radius as a planet with more iron and water, but less silicate. In order to understand the full range of solid exoplanets interior compositions, we examine the possibility of yet another ambiguity: coreless terrestrial exoplanets.

Differentiation in terrestrial planets is expected to include the formation of a metallic core. The terrestrial planets differentiated early in solar system evolution to have metallic iron cores and silicate mantles. Planetary differentiation is driven by density: metallic iron is denser than any mantle silicate mineral at any pressure. Its migration to the planetary core is proposed to require melting of the metal and likely of the silicate portion of the planet as well (Rubie et al. 2003; Terasaki et al. 2005; Jacobsen 2005). Heat is converted from the kinetic energy of accretion and from potential energy release during differentiation. The lower viscosity caused by heating and the gravity of a growing body both act to produce a hydrodynamically spherical planet; in a body over several hundred kilometers in diameter sphericity is expected and usually indicates differentiation.

Coreless planets have been discussed previously (Stevenson 1982, Valencia et al. 2006, 2007, Fortney et al. 2007, Seager et al. 2007, Sotin et al. 2007), as a possible end-member to the three broad categories of planetary materials: iron metal, silicate rock, and water. While these three components are each present in solar system terrestrial planets, the discovery of exoplanets enlarged the scope of prospective interior composition, motivating models of simple end members. Here, for the first time to our knowledge, we follow plausible accretion processes to demonstrate the plausibility of coreless terrestrial planet formation. We also present the observational ramifications of a coreless planetary structure.

# 2. THE CORELESS TERRESTRIAL PLANET HYPOTHESIS

We hypothesize that there are two accretionary paths to a coreless planet. In the first, the planet accretes from primitive material that was fully oxidized before accretion. This hypothesis may be supported by the existence of chondrite meteorites that have no metallic iron but bear excess water in the form of OH molecules bound into silicate mineral crystals. Such a planet may be more likely to form when temperatures are cooler, either later in accretion and/or farther from the star.

In the second hypothesis, the planet accretes from both water-rich (oxidized) and iron metal-rich (reduced) material. Instead of sinking as metal diapirs to form a core, the metal iron reacts with water to form iron oxide and release hydrogen. If this reaction is much faster than differentiation of a metal core, the iron is oxidized and trapped in the mantle, unable to form a core.

A critical basis for this second hypothesis for a coreless planet formation is that the oxidation rate of iron is faster than the sinking of iron to form an iron core. We now turn to a quantitative description of the rate competition between sinking iron droplets or solids and the rate of oxidation in order to understand under what conditions a coreless terrestrial planet may form under this second hypothesis.

Oxidizing metallic iron is a competition between the rate of oxidation and the rate at which the metallic iron sinks and is incorporated into the core of the planetesimal or planet, after which it is effectively shielded from the oxidizing agents. Core formation likely requires at least a partially molten planet (Jacobsen 2005). Oxidation likely also requires a liquid phase for highest efficiency. Here we will consider the conditions that allow oxidation of metallic iron in a partially- or fully-liquid magma ocean. An alternative scenario is one in which molten iron is percolating downward along grain boundaries through a solid silicate. The iron may be oxidized in this way if conditions in the solid are sufficiently oxidizing, but experiments have shown that liquid iron will not percolate through silicates at pressures greater than a few GPa (Terasaki et al. 2005).

Two first-order effects control the rate at which the metallic iron will sink toward the core: The size of the pieces of iron, and the effectiveness of convective entrainment of the pieces of iron in the fluid phase. The rate of sinking can be calculated as Stokes flow:

$$v_S = \frac{2(\rho_{solid} - \rho_{liquid})gr^2}{9\eta}, \qquad (1)$$

where $\rho$ is the density of the solid or liquid, as noted, $g$ is gravity, $r$ is the radius of the falling sphere, and $\eta$ is fluid viscosity. The density of the solid in this case is ~7,000 kg m$^{-3}$, and the density of the liquid is ~3,000 kg m$^{-3}$. Assuming a liquid viscosity of 1 Pas (Liebske et al 2005), a sphere of iron 1 cm in radius will sink with a velocity of about 0.002 to 0.20 m sec$^{-1}$ over a range of planetary gravities from 0.2 (for a body slightly smaller than Vesta) to 20 m sec$^{-2}$ (a super-Earth).

We consider a 500-km magma ocean: a small differentiated planetesimal may have a silicate mantle 500 km thick, and the Earth and larger bodies can obtain magma oceans of this depth from catastrophic impacts. Deeper magma oceans are hypothesized to have occurred following impacts on the scale of the Moon-forming impact. A deeper magma ocean, in these analyses, will have the effect of lengthening the sinking time of the droplets. If a liquid layer is 500 km deep, a centimetric drop will take between $10^6$ and $10^8$ seconds to reach the bottom, over the range of planetary sizes considered, if Stokes flow is a reasonable approximation. Iron drops smaller than a centimeter can take closer to $10^9$ seconds to sink in a small body (tens of years), and drops 10 cm radius take about $10^4$ seconds to reach the bottom of a large body (hours).

The Stokes sinking time is an estimate of the minimum time period available for oxidation. Convective velocities in magma oceans may be very high (Solomatov 2000; Siggia 1994) and sufficient convective forces may exist to delay the persistent sinking of the dense metal spheres. Solomatov et al. (1993) suggested that particles smaller than diameter $D$ with density difference $\Delta\rho$ between themselves and the magma ocean liquids would remain entrained in a magma ocean with viscosity $\eta$ and surface heat flux $F$, as given by

$$D \approx \frac{10}{\Delta\rho\, g}\left(\frac{\alpha \eta g F}{C_p}\right)^{\frac{1}{2}}, \qquad (2)$$

scaled by thermal expansivity $\alpha$, gravity $g$, and heat capacity $C_p$. Magma ocean fluid viscosities are expected to be 1 Pas or less (Liebske et al. 2005). The high density of the metal means that entrainment is only possible by high-viscosity fluids. For a heat flux of $10^4$ W/m$^2$, even on the smallest bodies, fluid viscosity must reach $10^5$ Pas before metal spheres of centrimetric size may be entrained in the flow. Heat flux varies by only about two orders of magnitude during the period when sufficient liquid exists to allow the metal droplets to fall. For each order of magnitude heat flux falls, the radius of the droplet that can remain entrained falls by a factor of three. Here we consider centimeteric drops, that will not be entrained in flow unless viscosity changes by more than four orders of magnitude, which it will not do without forming a crystal network and preventing any further settling.

Viscosity of the magma ocean will change during solidification. The rate and degree of viscosity changes in cooling magma are not well understood, though increasing crystal fraction appears to have the largest effect on viscosity. On bodies as large as the Moon, crystals appear to sink efficiently and fail to form interconnected networks; this is evidenced by the flotation of the buoyant mineral plagioclase to form the anorthositic highlands on that body. Here we assume in larger bodies that viscosity remains low through the majority of solidification and will therefore not significantly alter settling time. Because the timescale of solidification of a planetary magma ocean is far longer than the Stokes settling time for metallic material (Abe 1997, Elkins-Tanton 2008), drops of centimetric size will therefore sink and will not be entrained in flow.

In small bodies such as planetesimals, low gravity may allow the formation of interconnected crystal networks at crystal fractions as low as 30 to 40% (Saar et al. 2001). This networked magma ocean will behave as a solid deforming through thermally-activated creep, and viscosity will rise by perhaps ten orders of magnitude or more. In the case of small bodies, therefore, we consider viscosity to be effectively constant during the first 30 to 40% of solidification, after which settling will be either significantly slowed, or prevented altogether. Small bodies with a solid conductive lid over their magma oceans solidify in millions of years (Weiss et al. 2008), and metallic droplets will therefore have time to sink before solidification proceeds past the 30 or 40%. Small bodies with a liquid magma ocean surface may solidify in less than 100 years, close to the timescale of settling. Centimetric drops of metal may therefore be trapped before they can finish sinking to the interior.

Therefore, metal drops larger than a fraction of a centimeter will fall at their Stokes velocities, unless the liquid contains a significant crystal fraction as it is likely to do only on small bodies such as planetesimals, and only if these small bodies have free liquid surfaces to allow rapid cooling. Minimum residence times for magma oceans 500 km deep in this analysis, in summary, vary from $10^4$ to $10^9$ seconds, over gravitational accelerations from 0.2 to 20 m sec$^{-2}$ and metal droplet sizes up to 10 cm in radius.

While the drop is sinking water or other oxidizing agents in the liquid are reacting with the metal. For the oxidizing reaction to move toward iron oxide, hydrogen (in the case of water as the oxidizing agent) must be able to diffuse away from the oxidizing iron. Hydrogen diffusion in silicate liquid is fast, so this process should not limit iron oxidizing.

Once the outermost layer of the metal drop is oxidized, however, the oxidation of the interior of the drop is dependent upon the diffusion time of oxygen into the metal. For a first-order calculation, over the temperature and pressure ranges of interest oxygen may be assumed to diffuse through metal at $10^{-8}$ m$^2$ sec$^{-1}$ (Sayadyaghoubi, Sun, & Jahanshahi 1995, Blundell, Reid, & Zapuskalov 2005). A simple complementary error function solution to the diffusion equation demonstrates that oxygen will diffuse about 1 cm in $10^3$ to $10^4$ seconds, similar to the minimum sinking times of centimetric metal droplets.

Therefore it is reasonable to say that in the absence of high viscosity fluids, metal drops can be no larger than a few centimeters in diameter if they are to oxidize fully before forming a core, but if they are centimetric or smaller and are sinking in an oxidizing liquid, they are likely to become oxidized.

Metal accreting to a planet is likely to come in the form of the cores of previously differentiated, unoxidized planetesimals. These cores must be broken apart into small particles for oxidation to go forward. For metal to be incorporated into a planet on the centimetric scale from previously differentiated planetesimals or embryos, the accretionary impact must be sufficiently violent to fully disaggregate the impactor, thought to be a reasonable likelihood (Rubie et al., 2003). If the metallic core of an embryo does not disaggregate but sinks rapidly into the interior of a growing planet (e.g. Canup 2004), it will avoid oxidation and form a metallic core.

Oxidation is far more likely on smaller bodies, where the high viscosity of cooling magma may trap sinking spheres before they can reach a core, or where sinking is sufficiently slow. These oxidized smaller bodies may then accrete into a planet. Oxidation on small bodies is more likely than on large bodies for a second reason, as well: High temperatures encourage reduction, and the formation of magma oceans on planets requires far higher temperatures than do planetesimals. The potentially cooler liquid conditions on planetesimals are therefore more likely to encourage oxidation.

## 3. MODELS

We now turn to planet interior models in order to investigate the internal density structure of coreless terrestrial exoplanets. We present a hierarchy of models, beginning with the most detailed model for a low-mass planet such as Mars, moving to a more generalized model for the interior structure of super Earths, and finally to a basic model to compare the radii for differentiated and coreless planets of the same mass, over a wide range of masses. The hierarchy of models enables us to explore the interior density structure evolution from a magma ocean to a coreless planet, to extrapolate to a range of masses, and to investigate the overall mass-radius relationships. Specifically, the level of detail used in the Mars model is possible in part because of data available on Mars compositions from meteorites, and from experimental results that indicate the temperature of the solidus, the likely mineralogical assemblages, etc. Because the planet has a relatively small pressure range, the thermochemical parameters for the minerals present in Mars' mantle have been thoroughly determined through experiments, Its resulting density profile comparisons are therefore highly robust and make a strong example. For hypothetical super-Earths, the simpler models are appropriate for purposes of comparison and prediction.

**3.1 Detailed Interior Density Structure Model**
This model predicts the interior composition and structure of a planet that solidifies from a liquid state.

The underlying assumption of the model is that bodies are accreted from planetesimals that originated at a range of orbital distances (e.g., Wiedenschilling 1977; Safronov & Vitjasev 1986; Kokubo & Ida 2000; Chambers 2006; O'Brien et al. 2006; Raymond et al. 2006). Because material from larger orbial radii is likely more water-rich, this scenario raises the possibility that rocky and metallic material from the inner solar system may be combined with water-rich material during the planet-forming process.

Planets are assumed to form from planetesimals created from materials similar to the chondritic meteorites from the collection of falls found on Earth. These meteorites represent primitive, undifferentiated material from the time of planetary accretion. Chondrites contain a variety of silicate components, including chondrules (perhaps the earliest condensed material from the planetary nebula), minerals such as olivine and pyroxene, silicate glass, and in some cases water- or carbon-rich veins, indicating alteration after formation.

A simple inverse relation exists between metal and water in chondrites and invites the description of a coreless terrestrial planet: the chondrites with water contents approaching zero (those with the least alteration) have the largest metallic iron fractions. Those with high water contents have little or no metals. High metal examples can contain metallic iron and nickel approaching 70 mass% (Benz et al. 1988; Hutchison 2004; Lauretta et al. 2007), while Wood (2005) reports up to 20 mass% of water in chondrites with little or no metals (Figure 1 and Table 1).

We assume that the bulk silicate planet has melted from the combined effects of heat of accretion and, if there is a metallic core, the potential energy release of core formation. In the coreless case, melting is facilitated by the addition of water, which lowers its melting temperature (a well-documented process on Earth, e.g. Mysen & Boettcher 1975; Grove et al. 2006). The resulting planetary magma ocean solidifies from the bottom upward because of the high-pressure intersection of the well-mixed magma ocean adiabat and its solidus, which has a lower slope (Abe 1997; Solomatov 2000).

For the detailed model of the Mars-sized planet, the planetary interior is shown immediately after overturn to compositional gravitational stability. The onset of thermally-driven convection, such as exists in the Earth, is slow and mixing of the mantle may be incomplete over the time scale of billions of years (as evidenced by the compositions of Martian meteorites). On Earth, however, the mantle appears to be more homogeneously mixed than is Mars' mantle. For reasons that are not entirely understood, thermally-driven convection may be vigorous enough on some planets to mix the mantle to homogeneity, while on other planets, convection is sluggish enough to allow original silicate differentiation to remain on some scale.

**3.2 Calculations for the detailed Mars-sized planet**
We calculate mineral compositions in equilibrium with the magma ocean composition using experimental distribution coefficients and saturation limits. Rates of equilibrium crystal growth are generally far faster than rates of solidification, so the assumption of equilibrium is reasonable. Some exoplanet interior models have assumed the planetary silicate composition consists purely of the mineral $MgSiO_3$ (e.g., Seager et al. 2007) while other have considered the addition of iron into the mineral structure of mantle minerals (e.g., Valencia et al. 2007; Sotin et al. 2007). Here we approach the complexity of mineral assemblages believed to exist in the Earth and Mars, and by extension in larger planets, by considering the silicate to consist of crystalline mineral components that include not just $MgO$ and $SiO_2$, but also $FeO$, $CaO$, and $Al_2O_3$. Together these five oxides are thought to make up >99 mass% of the Earth's mantle ($MgO$ and $SiO_2$ together make up ~84%). More importantly, we calculate compositions and densities for a series of mineral assemblages that are likely to exist at equilibrium at planetary interior pressures. Perovskite, the volumetrically-dominant mineral in the Earth's mantle, is stable from ~24 GPa to ~110 GPa, and is therefore not the dominant mineral in all planetary mantles.

In these models, at pressures greater than 24 GPa, perovskite and magnesiowustite are assumed to crystallize. From 22 to 14, γ-olivine (ringwoodite) and majorite are stable; from 14 GPa to 2 GPa garnet + clinopyroxene + orthopyroxene + olivine crystallize; from 2 GPa to 1 GPa spinel + clinopyroxene + orthopyroxene + olivine crystallize, and from 1 GPa to the planetary surface the phase assemblage is plagioclase + clinopyroxene + orthopyroxene + olivine (Figure 2). We calculate densities using a Birch-Murnaghan equation of state with parameters gathered from the experimental literature. For further details and a complete listing of parameters, see Elkins-Tanton et al. (2003), Elkins-Tanton & Parmentier (2005), and Elkins-Tanton (in review).

Solidification produces a density profile in the planet that is compositionally unstable to gravitational overturn. Compositional instability is produced by iron enrichment during progressive solidification from the bottom of the mantle upward. At the relatively low viscosities of the hot, solid silicate planet (perhaps in the range of $10^{17} – 10^{19}$ Pas), the gravitationally unstable material can flow in the solid state to gravitational stability, with higher-iron compositions at lower planetary radii. This critical step in planetary evolution from a magma ocean stage results in a mantle that is gravitationally stable, with densest silicates lying at lowest radii, and most buoyant silicates nearest the planetary surface. Thus planetary differentiation includes not only separation of metallic from silicate stages, but also compositional separation of silicates.

Though this model for planetary magma ocean solidification and overturn to stability involves physical and chemical simplifications, it has been validated against both experimental results for compositions and densities of minerals, and against evidence for early formation processes on Mars and the Moon (Elkins-Tanton et al. 2003, 2005a, b; Elkins-Tanton 2008; see these papers for full details of the calculations presented here).

We begin with two planetary compositions. The first assumes a planet with a metallic core fraction and a silicate mantle based on an average CH chondrite composition, from which the metallic iron is sequestered in a core and the remaining composition makes up the mantle (Table 2). The second, coreless case assumes the bulk composition of the same average CH chondrite, but with all the metals oxidized by addition of water, and incorporated into a silicate bulk composition (Table 2). The CH chondrite is chosen because it has a relatively high fraction of metallic iron (Figure 1). These compositions are applied to a planet with radius 3,400 km, about the size of Mars.

To enable comparisons between model results and solar system bodies, we quantify the distribution of mass within bodies using the planet's moment of inertia. For comparison, the moment of inertia of the planets resulting from these models can be calculated as

$$I = \int_0^{r_f} (\rho(r)) r^2 dr, \quad (3)$$

taking the density with radius $\rho(r)$ from the model results. The planet's moment of inertia factor $K$, making comparisons among planets possible, is the moment of inertia is divided by the mass of the planet $M$ and its radius $R$ squared:

$$K = \frac{I}{MR^2}. \quad (4)$$

A homogeneous sphere has a moment of inertia factor of 0.4, and the moment of inertia factor decreases if density increases with depth.

## 3.3 General Interior Density Structure Model

This model predicts the interior structure of a planet based on an assumed homogenous composition. Unlike our first more detailed model, this model does not calculate solidification from a liquid but starts with the assumed homogeneous composition. The model does include mineral phase changes, pressure, and temperature which all affect density.

To investigate the effects of iron on density profiles of planets larger than Earth, we use the concept of an Mg#, where the Mg# of a silicate material is its molar Mg/(Mg+Fe). This ratio is particularly useful because the majority of silicate minerals in planetary mantles have crystal lattice sites that accept either Mg or Fe; by adding iron, magnesium is excluded and the mineral becomes denser. In other words, if the planet is highly oxidized and iron that would otherwise be in a metallic core has mixed into the silicate mantle, the planet's silicate will be far denser than it was without the additional iron.

Earth's mantle is highly magnesian, and a typical terrestrial upper-mantle olivine mineral grain has an Mg# of ~0.93. We investigate planets with silicate compositions that would create upper-mantle olivines with Mg#s of 0.2 (high iron) and 0.9 (low iron). We assume the silicate portion of the planet is well-mixed, and adjust the Mg#s of each mineral phase in equilibrium with the olivine Mg# by amounts consistent with experimental data. We assume that the mineral assemblages described in the section above are the stable assemblages (Figure 2), and in addition at pressures greater than 110 GPa the mineral assemblage is assumed to be pure post-perovskite phase. Though there are likely additional phase changes at higher pressures, they have not been characterized in experiments, so our highest-pressure phase is post-perovskite.

We calculate densities for the mineral assemblages using a Burch-Murnaghan equation of state, as described above, using separate parameters for the iron- and magnesium-bearing fractions of each mineral phase. The temperature profile of the planet begins at 50°C at the surface and increases along a linear conductive geotherm through a 150 km-thick conductive lithospheric lid, at the bottom of which it bends into an adiabat through a well-mixed silicate interior with a potential temperature of 1,300°C. The plate thickness is arbitrary and unimportant in modeling results because the temperature change across such a plate causes negligible density changes in the resulting model; it is included as an approximation of a terrestrial planet structure

Each of these models ends at what would be approximately half the radius of the planet. If the planet has an underlying metallic core, densities will jump to far higher values. If instead the planet consists entirely of silicate material, the density profile shown would extend to the planet's center. The gravity with depth relation in a planet with a metallic core would differ from that of a coreless planet of identical mass and bulk composition, but the effect of these gravitational differences on these density profiles would be small compared to the compositional density difference, and are neglected here.

### 3.4 Mass-Radius Model
To investigate a broad range of planet masses and radii we use the model described in Seager et al. (2007). This model solves the equations of mass of a spherical shell and of hydrostatic equilibrium, using a prescribed equation of state. The boundary conditions are zero pressure at the planet surface (where the pressure is equal to 0) and at the planet center a zero mass as well as central pressure chosen for the planet interior. Gravity is self-consistently included. These conditions set the total mass and radius of the planet. The equations of state used are for Fe (epsilon) (Anderson et al. 2001b), $MgSiO_3$ perovskite (Karki et al. 2000) $Mg_{0.88}Fe_{0.12}SiO_3$ perovskite (Knittle & Jeanloz 1987) and $Mg_XFe_YSiO_3$ perovskite from (Elkins-Tanton in review). Although we consider the planet to be a uniform temperature throughout and have omitted to consider phase changes, we have previously shown that these assumptions are appropriate for understanding the mass radius relationships for super Earth exoplanets (but not the interior density structure) (Seager et al. 2007).

### 4. RESULTS
### 4.1 Mantle density profiles and moments of inertia for a Mars-sized planet
Density as a function of radius is the most useful result of these models, because it emphasizes the difference in interior structure between coreless and cored planets. Results for the Mars-sized planet are given in Figure 3. The mantle composition with a metallic core produces a planet with a dense core and a low-density mantle. The metallic core has a density on the order of 7,500 kg m$^{-3}$. The silicate mantle's density averages about 5,000 km m$^{-3}$ less than the core, and density range within the silicate mantle itself is only 400 km m$^{-3}$. The fully-oxidized, coreless planet has a density range within only silicates of about 700 km m$^{-3}$ from the planet's center to its surface. Thus with approximately the same bulk compositions, the two planets produce vastly different density profiles.

A coreless silicate planet will have an average uncompressed density higher than a similarly composed planet with a core, and its moment of inertia factor will be higher. In the examples here, the coreless planet has an average uncompressed density of 3,700 km m$^{-3}$, while the planet with a metallic core has an average uncompressed silicate mantle density of only 3,370 km m$^{-3}$.

Near their surfaces, the coreless planet is denser by ~10%, while at a depth of about half their radius the coreless planet is denser by ~20%. At depths that reach its core, the planet with a core is 200% denser than the coreless planet. Each ends with a gravitationally stable silicate mantle, though the coreless planet has a far more stable silicate density gradient, which is therefore far more resistant to onset of thermal convection.

The moment of inertia factor can be measured remotely for solar system planets and is therefore of use in discriminating internal structures and comparing among planets of different sizes and masses. The planet with a metallic core has a moment of inertia factor of 0.369 (the core is assumed to have the density profile for Mars from Bertka & Fei 1998), while the coreless planet's is 0.389, a significant difference. Because these moment of inertia factors are normalized by planetary mass and radius, they are applicable to planets with similar interior compositions but different masses.

These model results are in good agreement with known moment of inertia factors: the Moon's moment of inertia factor is 0.392, indicating a small or nonexistent metallic core. The Moon's silicate interior is known to be differentiated. The next largest moment of inertia factor of the terrestrial planets is Mars', at 0.366 (Folkner et al. 1997). Earth's, Venus', and Mercury's moment of inertia factors are all near 0.33, Io's 0.377 (Anderson et al. 2001a) and Callisto's 0.359 (Anderson et al. 1998).

**4.2 Density profiles in super Earths**
The density profiles for larger planets shown in Figure 4 clearly demonstrate the significance of a high iron content in the silicate planetary material. The denser profile at each mass represents a planet with silicate mantle of Mg# ~0.2, and the lower density profile of Mg# ~0.9.

Steps in the profiles represent changes in the mineral phase assemblages. Each mineral, and in fact each compositional end-member of each mineral (for example, $MgSiO_3$ perovskite vs. $FeSiO_3$ perovskite) has its own response to pressure and temperature, creating differences in density even with the same bulk composition. The density profiles diverge with depth for this reason: for most mantle silicate minerals, the iron-bearing end-member has a larger bulk modulus than the magnesian end-member.

At the bottom of each profile (representing about half the radius of the planet), the high-iron profile is ~20% denser than the low-iron profile. The radii in these models are approximate; the effects of these compositional changes on planetary radius will be discussed in the following section.

**4.3 Radius comparison for coreless and differentiated exoplanets of the same mass**
Most investigations of exoplanet interior composition has heretofore assumed that the exoplanets are fully differentiated with an iron core (we will call this a cored planet; see Valencia et al. 2006). In order to assess potential misinterpretation we plot radius curves for a coreless and cored planet of the same total mass for the mass range 1 to 10 Earth masses (Figure 5). In comparing a coreless and cored planet, we consider planets with the same total iron mass fraction. For our first example we compute a cored planet with a 30% iron core by mass fraction and a 70% by mass silicate perovskite mantle of $(Mg_{0.88}, Fe_{0.12}SiO_3)$; approximating the Earth's core/mantle mass division. A coreless planet with the same total iron mass fraction but consisting of only a perovskite mantle would be composed of $Mg_{0.24}Fe_{0.76}SiO_3$. We note that this form of perovskite is highly iron-rich and that this helps to illustrate how high Earth's bulk iron content actually is.

Our second example is for a cored planet with a 10% iron core by mass and a 90% by mass silicate mantle of $(Mg_{0.88},Fe_{0.12}SiO_3)$. The corresponding coreless planet with the same iron mass fraction is composed of $Mg_{0.7}Fe_{0.3}SiO_3$.

For the same iron mass fraction the cored planet has a slightly higher average density than the coreless planet. The higher average density is foremost caused by the higher zero-pressure density of iron (8.1 g cm$^{-3}$) as compared to perovskite (~4.3 g cm$^{-3}$), and second by the compression of iron in the core. This higher average density results in a smaller radius for the same total mass, as shown in Figure 5. The radius of a cored planet is only a few percent smaller than the corresponding coreless planet, a value smaller than anticipated radius measurement uncertainties. This indicates a new major source of uncertainty in interpreting exoplanet radii. One of the major conclusions in this paper is that the interior structure of an exoplanets cannot be inferred from a mass-radius measurement alone.

## 5. DISCUSSION AND SUMMARY

We hypothesize that there are two accretionary paths to a coreless planet. In the first, the planet accretes from material that was fully oxidized before accretion. This hypothesis may be supported by the existence of chondrite meteorites that have no metallic iron but bear excess water (e.g., chondrite classes CI, CM, CR, CO, CV, and CK; see Figure 1 and Table 1). Such a planet may be more likely to form later in accretion, when temperatures in the planetary nebula have fallen, and may be more likely to form farther from the star where volatile-rich material is more common (Machida & Abe 2008).

In the second hypothesis, the planet accretes from both oxidized and reduced material (such as the reduced material of the chondrite classes CH, H, EH, EL, and CB; see Figure 1 and Table 1), and oxidation of metals occurs in a well-mixed magma ocean or partially-molten slurry. The current picture of planet formation is the accretion of differentiated planetesimals and planetary embryos ranging in size from tens to thousands of km in radius. These accreting small bodies may therefore have metallic iron cores. We have, however, shown that iron droplets must be centrimetric or smaller to oxidize before they sink to form a core. Accretion may be energetic enough to fully melt and mix all materials both in the impactor and in the impact site. The likelihood of the disaggregation or melting of the impactor is not well understood by models. Clearly this depends on the relative sizes of the impactor and the growing planet and of the velocity and angle of impact.

When iron is oxidized by water, hydrogen is released. Under the first hypothesis, in which oxidation occurs before accretion, the resulting planet will have no hydrogen atmosphere because the hydrogen reaction products will have been lost to space from the pre-accretion materials. Under the second hypothesis, however, hydrogen is released from the growing planet as oxidation progresses, and if the planet is sufficiently massive the degassed hydrogen atmosphere can be retained. Using the average chondrite class compositions in Table 2 as simple illustrative end-members, complete iron oxidation for metallic iron-rich compositions can require up to approximately 20 mass% of the planet in additional oxygen. If volatile-rich comet-like solar system material is roughly half water, then to fully oxidize a planet that begins with the highest fraction of iron metal requires the final planet to consist of roughly 40% volatile-rich added material. Planets may be expected, therefore, to attain varying degrees of oxidation depending upon the oxidation states of their accreting material and the degree of mixing the accreting material experiences.

A fully silicate coreless planet will not have a liquid core and therefore no magnetic dynamo. Earth's magnetic field is critical in shielding the atmosphere from energetic particles.

Beginning accretion with material already oxidized, or oxidizing material during accretion by addition of volatile-rich material, imply the possibility of planets with a full range of metallic core masses, from zero to the maximum metallic iron available (which in the solar system is currently thought to be the CB chondrite, with as much as ~70 mass% metallic iron [Lauretta et al. 2007], equivalent to the core mass of Mercury). If the most reducing conditions existed closest to the Sun and the most volatile-rich at increasing radius from the Sun, then the terrestrial planets' core masses might be expected to follow this trend, which in fact they do. Mercury, closest to the Sun, has a core equal to 60 to 70 mass% of its planetary mass, while the Earth's is ~32 mass% of the planet, Venus' ~25 mass%, and Mars', only 15 – 20 mass%. This increase of silicate mass of the planet with radius supports the possibility of a coreless silicate planet at larger orbital radii.

The addition of volatiles may occur preferentially at greater radii, and it may occur preferentially at later times during accretion. The large metallic cores of the terrestrial planets require that early accretion occurred in a highly reducing environment. The oxygen fugacity (reactivity and availability of oxygen) in the solar nebula has been estimated to be as low as IW-5 (five log units below the iron-wüstite buffer, or very highly reducing) (Jurewicz 1995). The Earth and Mars (and perhaps Venus) later obtained a significant water content, possibly by later accretion of hydrous materials (Ringwood 1979; Wänke 1981; Hunten et al. 1987; Dreibus et al 1997), or possibly from water adsorbed from the planetary nebula onto small, undifferentiated accreting materials (Drake et al. 2004; Righter 2004; Drake & Campins 2006). The high oxygen fugacity of the Earth's mantle presents a problem when compared to the hypothetical reducing planetary nebula: Righter (2004) proposed that a volume of water equivalent to 50 Earth oceans is required to raise the mantle from IW-3 to its present IW+3.5. Had this oxidation occurred earlier in formation, the Earth may not have retained its large core.

The sequence of reducing conditions that allow formation of a metallic core, followed by addition of more water- and carbon-rich material that does not interact with the metallic core, is the most consistent scenario for formation of Venus, the Earth, and Mars. They are known to contain metallic cores, silicate mantles, and some volatile content. Mercury, with its large metallic core and apparent lack of water, may have only experienced the first stage of accretion, under extremely reducing conditions. Terrestrial planets, therefore, appear to have formed under either reducing conditions, or reducing conditions followed by oxidizing conditions.

The outer planets, in contrast, are unlikely to have experienced reducing conditions because of the high volatile content of the outer disk. Though none has been definitively identified, it is possible that one or more of the outer planets or their moons contain a fully oxidized central silicate core with no metallic iron interior. Both timing and original radius in the solar system, therefore, may control the oxidation state of a planet. Highly oxidized silicate planets may attain sphericity and complete differentiation, but lack any metallic core. Since Calllisto is believed to be a mixture of rock and ice without a metallic iron core, it may be the result of mixing high volatile content materials with rocky materials and in the process oxidizing all of the metallic iron.

 We have described two pathways to forming coreless terrestrial planets, implying that terrestrial planets with a range of core sizes are possible. Our interior models show that the radii of coreless and cored are similar within a few percent, and that this will lead to ambiguities in interpreting the bulk interior composition based on measured masses and radii. Our interior models also show that the interior density structure of a coreless and cored planet are very different and lead to measurably different moment of inertia factors.

## 6. ACKNOWLEDGMENTS

## 8. FIGURES

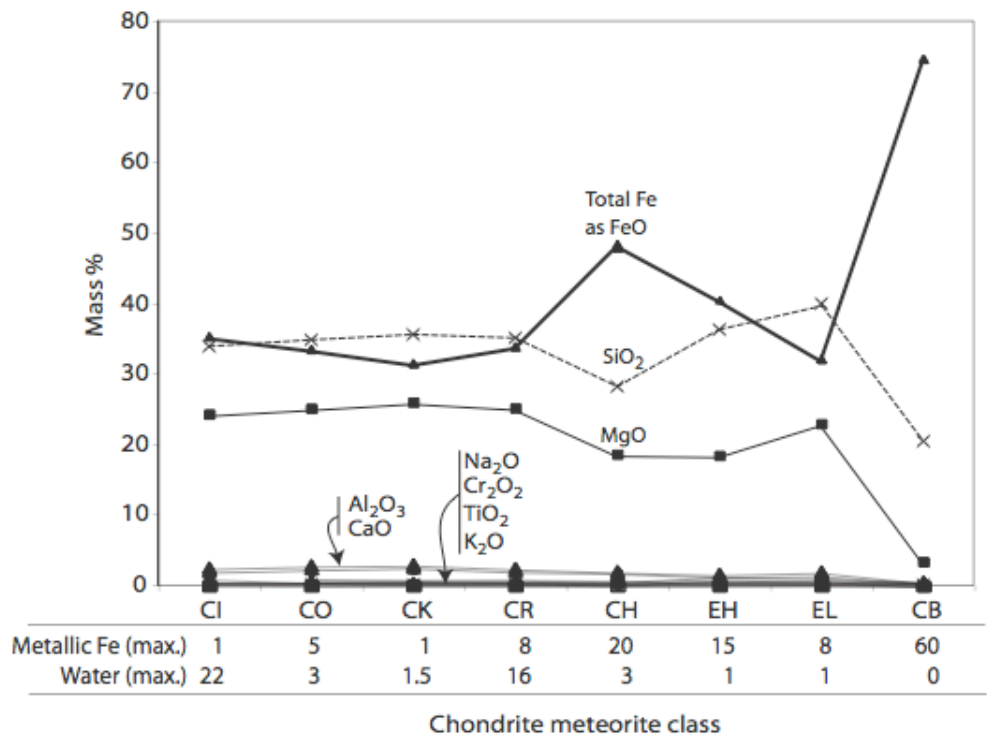

**Figure 1.** Average bulk compositions of chondrite meteorite classes, with total iron expressed as FeO. Metallic iron and water maxima in mass % are given below, demonstrating their rough negative correlation. The CH bulk composition is used for detailed models in this paper. Data from Hutchison (2004) and Lodders & Fegley (1998).

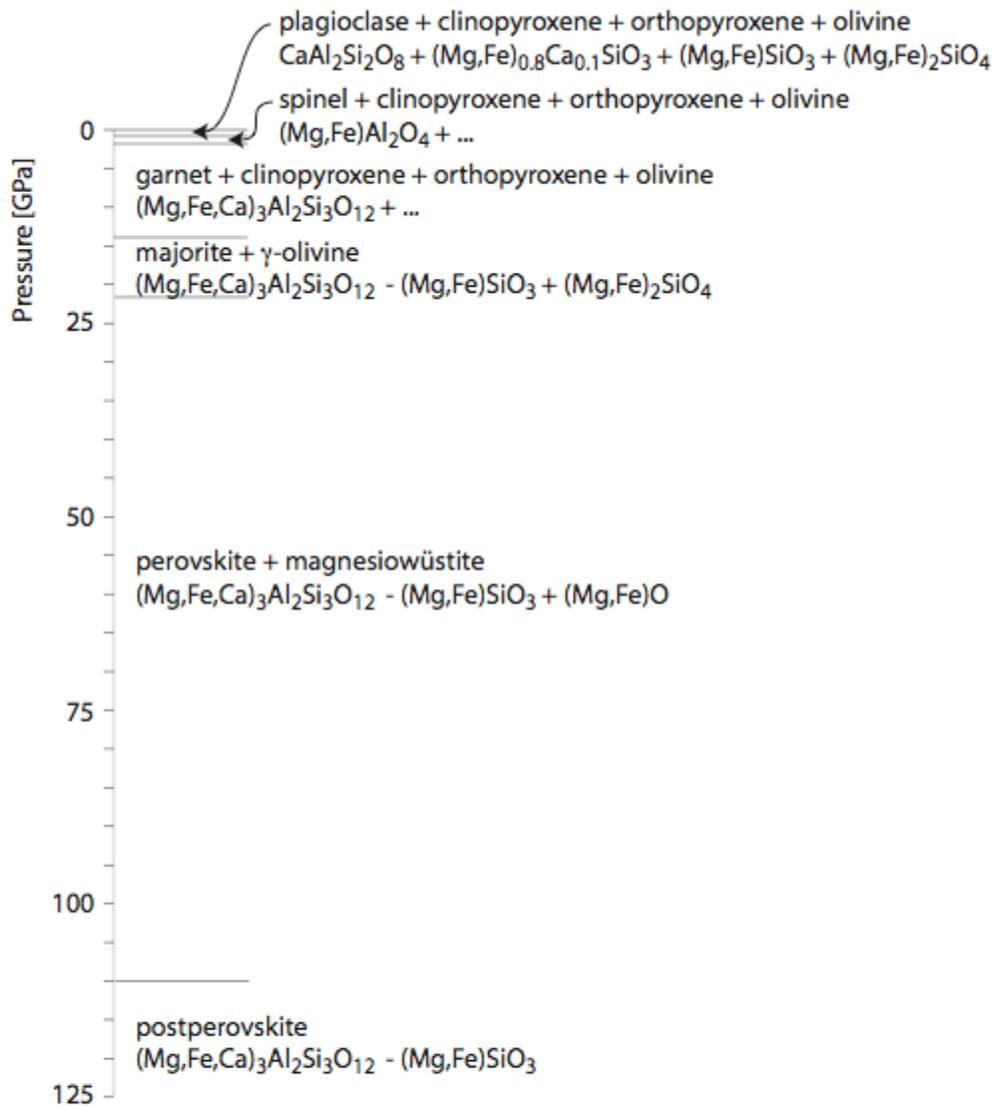

**Figure 2.** Stable mineral phase assemblages for silicates of bulk composition similar to Earth's mantle. The percentages of the phases at each pressure range depends upon bulk composition. These phase assemblages are used in the Mars-sized detailed model, and in the multiple Earth-mass calculations of density profiles.

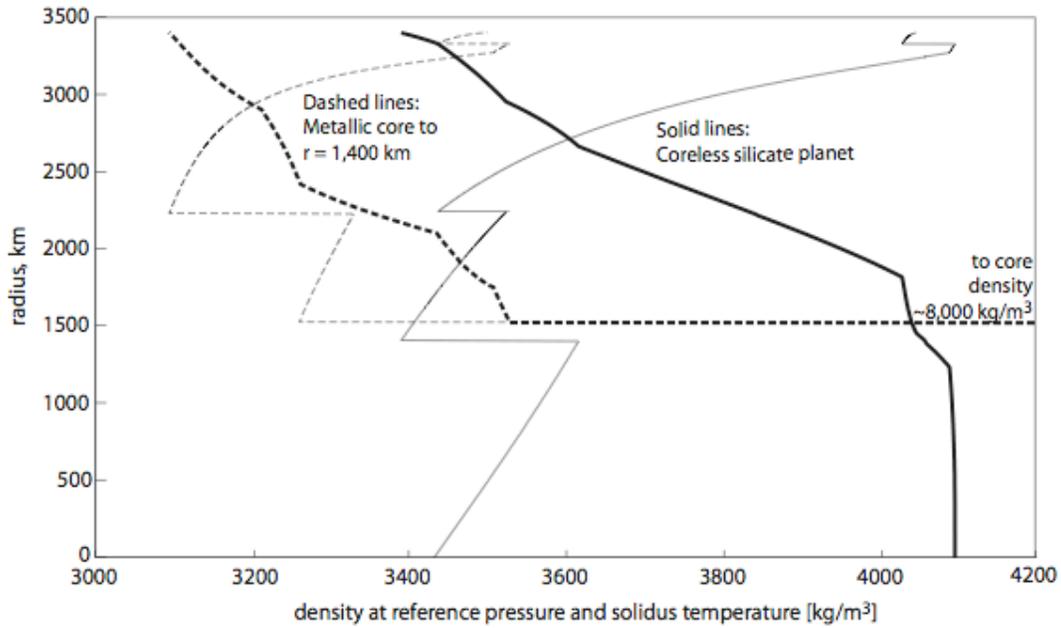

**Figure 3**. Density as a function of radius for two planetary models. Dashed lines: Metallic core to ~50% of planetary radius, and then silicate mantle composition (Table 1). Solid lines: Average CH chondrite composition, fully oxidized, over entire planetary radius. Fine lines indicate the unstable density gradient of silicates differentiated in magma ocean solidification, and bold lines indicate the stable silicate planet following overturn of the magma ocean differentiates in the solid state. The oxidized composition is ~10% denser in the upper half of the planet's radius, and ~40% less dense in the inner half when compared to the metallic core of the other planet.

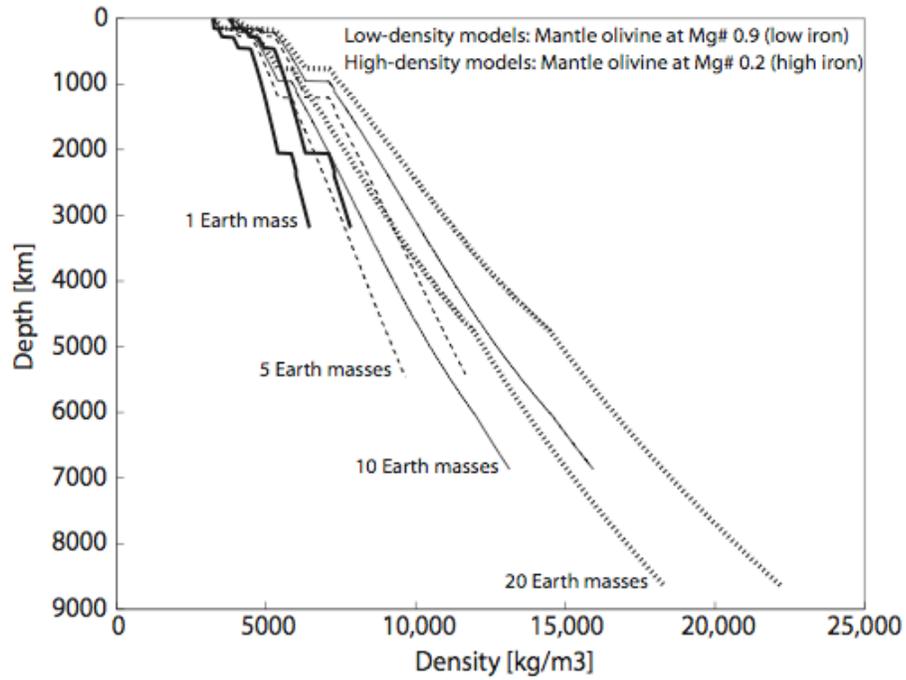

**Figure 4.** Density profiles in high-iron (Mg# = 0.2) and low-iron (Mg# = 0.9) well-mixed solid planetary silicate interiors. In each case a density profile to approximately half the planetary radius is shown.

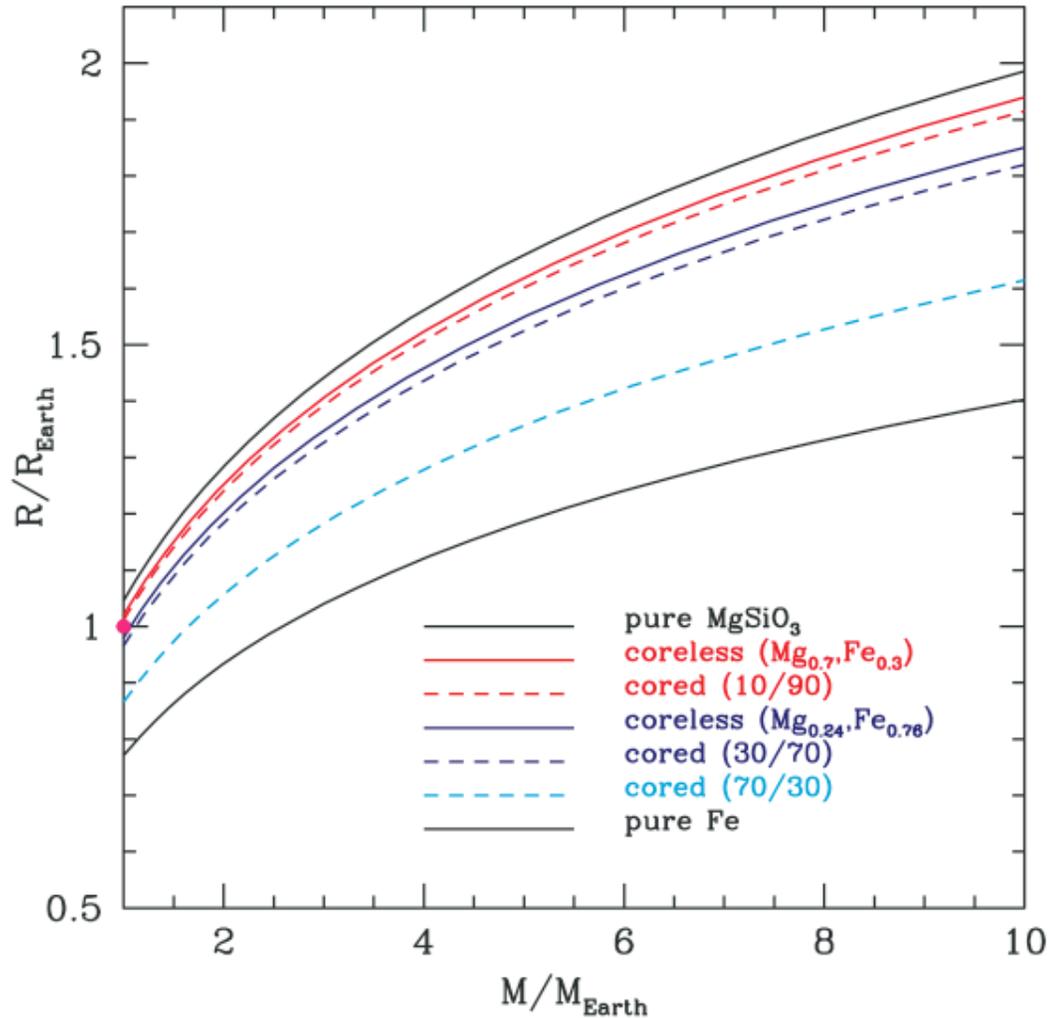

**Figure 5.** Comparison of mass-radius relationships for cored and coreless exoplanets. The top and bottom curves are homogeneous planets for illustration, composed of pure MgSiO3 and pure Fe respectively. The red curves correspond to a coreless (solid curve) and cored (dashed curve) planet with a 30% by mass metallic iron core and a total iron mass fraction of 34% (i.e., including the iron in the mantle). The blue curves correspond to a coreless (solid curve) and cored (dashed curve) planet with total iron mass fraction of 76%. The cyan dashed curve is a cored planet with 72% total mass fraction; there is no corresponding perovskite mineral that can contain the same iron mass fraction.

# 9. TABLES

**Table 1.** Average compositions of chondrite meteorite classes when fully oxidized. Initial compositions from Lodders & Fegley (1998), Lauretta et al. (2007), and Jarosewich (1990).

| chondrite class | | CI | CO | CK | CR | CH | EH | EL | CB |
|---|---|---|---|---|---|---|---|---|---|
| volatiles | C | 4.3 | 0.4 | 0.2 | 2.1 | 0.8 | 0.4 | 0.4 | 0.0 |
| | S | 13.4 | 4.4 | 3.5 | 3.9 | 0.7 | 10.3 | 5.8 | 0.3 |
| silicates | $Na_2O$ | 0.8 | 0.6 | 0.4 | 0.5 | 0.2 | 0.8 | 0.7 | 0.1 |
| | MgO | 19.9 | 23.8 | 25.0 | 23.6 | 18.3 | 16.4 | 21.5 | 3.3 |
| | $Al_2O_3$ | 2.0 | 2.6 | 2.8 | 2.3 | 1.9 | 1.4 | 1.8 | 0.4 |
| | $SiO_2$ | 28.2 | 33.5 | 34.6 | 33.3 | 28.1 | 32.7 | 37.8 | 20.8 |
| | $K_2O$ | 0.1 | 0.0 | 0.0 | 0.0 | 0.0 | 0.1 | 0.1 | 0.0 |
| | CaO | 1.6 | 2.2 | 2.4 | 1.9 | 1.8 | 1.1 | 1.3 | 0.4 |
| | $TiO_2$ | 0.1 | 0.1 | 0.2 | 0.1 | 0.1 | 0.1 | 0.1 | 0.0 |
| | $Cr_2O_3$ | 0.5 | 0.5 | 0.5 | 0.5 | 0.4 | 0.4 | 0.4 | 0.3 |
| | FeO | 29.0 | 31.9 | 30.3 | 31.8 | 47.6 | 36.2 | 30.0 | 74.4 |
| | total | 100.0 | 100.0 | 100.0 | 100.0 | 100.0 | 100.0 | 100.0 | 100.0 |
| oxygen excess or defecit after oxidizing all iron, as a fraction of planetary mass | | 15.9 | -2.4 | 0.2 | 1.6 | -6.3 | -15.5 | -12.6 | -19.6 |

**Table 2.** Silicate bulk compositions (mass %) used in detailed models of the Mars-sized planet. The two compositions are based on an average CH chondrite meteorite composition (Lodders & Fegley, 1998).

|  | Planet with metallic core | Coreless planet |
| --- | --- | --- |
|  | Metallic iron (24 mass%) removed into a core; remaining composition (shown here) used to make a silicate mantle | Metallic iron oxidized and complete bulk composition (shown here) used to make a coreless silicate planet |
| MgO | 27.1 | 18.7 |
| $Al_2O_3$ | 2.9 | 2.0 |
| $SiO_2$ | 41.8 | 28.8 |
| CaO | 2.6 | 1.8 |
| FeO | 25.5 | 48.7 |
| total | 100.0 | 100.0 |
| Mg# | 0.65 | 0.41 |